# Data Analysis on the High-Frequency Pollution Data Collected in India

Lamling Venus Shum, Manik Gupta, Pachamuthu Rajalakshmi
v.shum@ucl.ac.uk, manik.gupta@eecs.qmul.ac.uk, raji@iith.ac.in
University College London, Queen Mary University of London, Indian Institute of Technology, Hyderabad

**ABSTRACT**

Fine grained 1Hz Carbon Monoxide pollution data were collected on a busy road in Hyderabad, India. In this paper we report the findings from analysing the experimental data, in which it was found that the data were log-normally distributed and nonlinear. The dominant frequencies at peak hours were caused by the pattern of traffic flow.

## 1. INTRODUCTION

The impact of pollution is of significant scientific, social and economic interest to countries across the globe [1]. In developed countries, in which considerable control is exercised over emissions, air pollution from traffic constitutes one of the most significant remaining sources of exposure for individuals. In countries with rapidly developing economies, the rapid, uncontrolled, rise in traffic has led to a corresponding decrease in air quality for many in urban environments. In both cases, such pollution affects not only the general quality of the environment, but it also has direct and significant effects on the health of the population [1].

Our research aim is to examine the fine structure of pollutant distribution within defined urban environments. At present, such measurements are taken by using relatively costly monitoring stations at predetermined locations. In our work, a wireless sensor network is used as an alternative to the traditional approach, in which a significant number of low cost sensor nodes are deployed along the street at different heights. These sensors take regular readings at all locations, allowing the reconstruction of pollutant flow and the density profile of pollutants across the three dimensional space.

The ongoing research that forms the basis of this paper is a collaborative effort between teams in the UK and in India [2][4][10]. We have collected several sets of pollution data in both countries using our Orisen Carbon Monoxide monitors with high temporal and spatial resolution in relation to the norm for atmospheric science. Earlier work focus on developing and calibrating sensor devices to collect fine-grained, accurate data for scientific study and algorithm development [2]. The present phase of the design focuses on achieving energy savings in the devices for longer deployment.

The paper reports some data analysis results from a pollution measurement experiment conducted in Kukatpally, Hyderabad, India in February 2012.

## 2. DATA CHARACTERISTICS

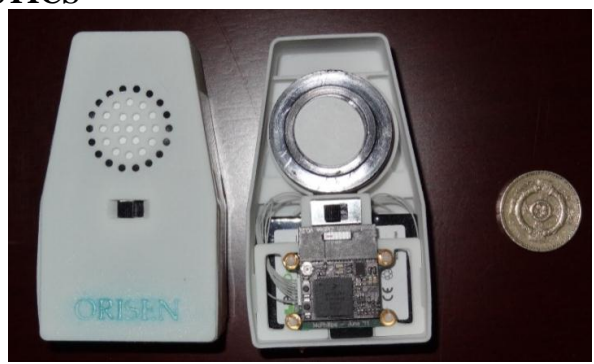

**Figure 1: Orisen CO monitors used in the experiment**

An experiment was conducted near Hyderabad Central Mall at the junction between Punjagutta Road and Nagarjuna Circle Road, Hyderabad, India on 9-10 February 2012 in order to collect pollution data including Carbon Monoxide (CO), Carbon Dioxide (CO2) and particular matter (PM10) for 36 hours. Orisen pollution monitors, as shown in Figure 1, previously known as the Bracelet wireless system [3][4], were developed and deployed in experiments carried out in the UK, India and Cyprus. Each device is equipped with temperature, humidity (Sensirion SHT25) and Electrochemical carbon monoxide sensors (KWJ Engineering Ltd RCO100F).

The CO data are collected at 1 sample/s and are logged as 12-bit ADC values. The values represent the voltages measured by the CO signal conditioning board described in [3]. The raw ADC values must be converted into part-per-million (ppm) for analysis by environmental engineers and this is done from individual calibration curves produced in a laboratory environment. Information on the design of the devices and calibration results against temperature and concentration can be found in [3][5].

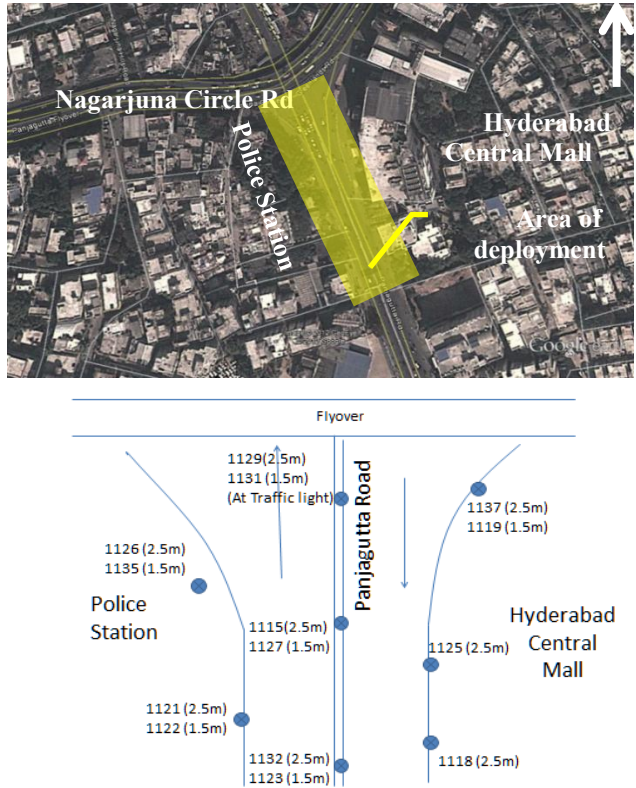

**Figure 2: Area of deployment**

## 2.1 Time Series and Distribution

Figure 3 and Figure 4 show plots of a section of time series recorded in the day and night time of the experiment.

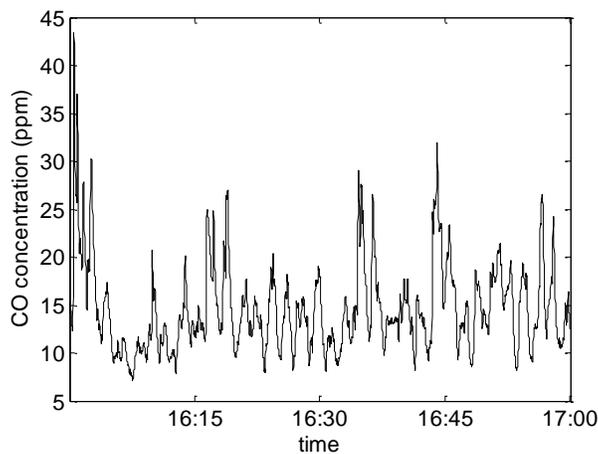

**Figure 3: Time series of CO concentration during 4-5pm on 9/2/2012, collected at Kukatpally, Hyderabad, India**

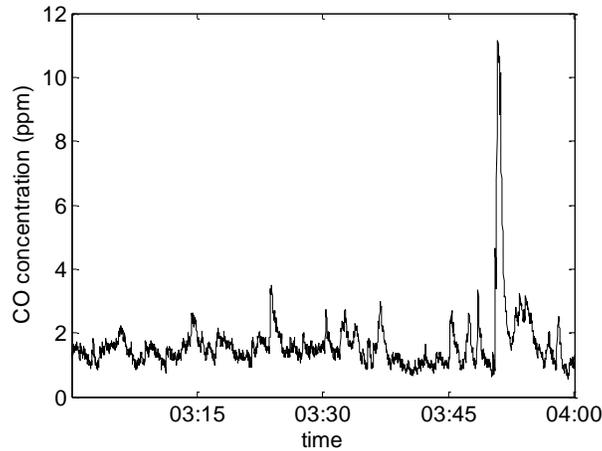

**Figure 4: Time series of CO concentration during 3am-4am on 9/2/2012, collected at Kukatpally, Hyderabad, India**

Earlier work, including [5][6], showed that pollutants were *log-normally distributed* at all time scales and appropriate averaging time and frequency were proposed. Some of the results have been used as guidelines for deriving pollution limits and standards. In this section, we attempt to study the data distribution during different time period and traffic conditions and suggest suitable models for the data.

In [5], the authors studied the data distribution and averaging time between 5 mins and 1 year and concluded that the averaging time had no effect on the sample distribution, which remained lognormal. With our high frequency Indian data, this lognormal distribution is observed during the day and night time with low to high traffic flow, as shown in Figure 5 and Figure 6. The time series in Figure 3 and Figure 4 appear to have some local trends and cycles, which are spurious and disappear after some time. According to [7] this is a typical feature of long-memory process

The Log-normal distribution is one of the heavy-tailed distributions. Both the frequency of high-valued events and their duration is small compared to low-valued event. Averaging the series smooth away the extreme and the rare occurrences of high CO values (>50ppm) have little effect on the hourly averages.

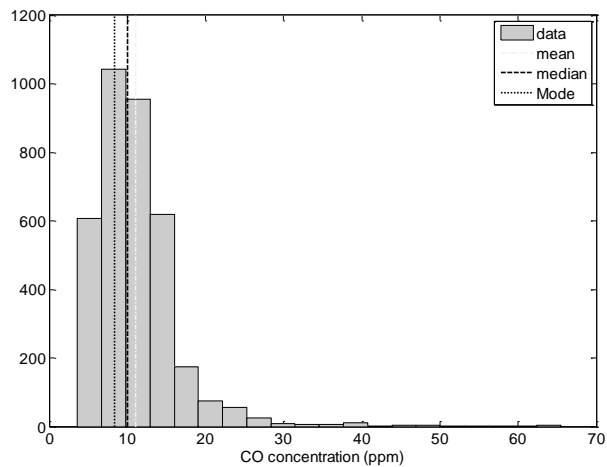

**Figure 5: Histogram of CO distribution during 4-5pm on 9/2/2012**

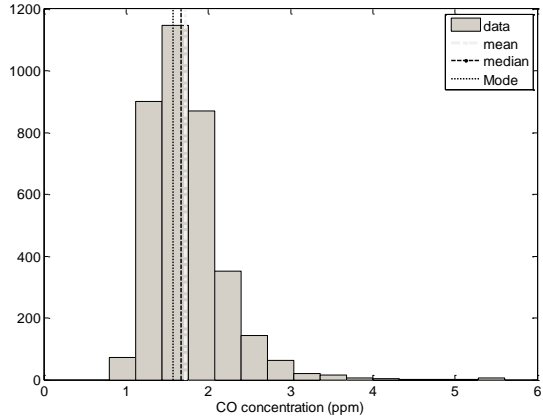

**Figure 6: Histogram of CO distribution at 3am-4am**

Intuitively, the variation of data during the day time is expected to be much greater than at night time and this matches our data as shown in Figure 3 and Figure 4. Consequently, more of the sampling budget must be allocated to day time statistics to achieve the same confidence level on the sample mean.

## 2.2 Autocorrelation and Nonlinearity

Autocorrelation (acf) and partial correlation (pacf) analysis were performed directly on the 1Hz data. The result of acf is shown in Figure 3, which shows a slowly decaying curve in which correlation only becomes insignificant after 40+ lags. pacf shows a very strong lag-1 correlation of 0.99 and lag-2 correlation of -0.8. Some literature would suggest this is indicative of an integrated ARIMA($p, d, q$) with $d \geq 1$, which would suggest that the first difference series is stationary.

However, the strong lag-1 correlation is due to the similarity of data within short time scales (<10s). The underlying dynamic is easily marred by the noise[1] in an outdoor environment. One way to observe the dynamic of pollution caused by traffic is to apply a low pass filter to eliminate high frequency noise. Alternatively, the dynamic can be observed more clearly by subsampling the pollution series at 0.1Hz and preserving the extremes. Figure 7 shows the acf of a subsampled 0.1Hz series. The correlation coefficients first decay to an insignificant level at around lag 5 (50 seconds). A clear oscillation is observed in the figure, which will be explained later. During the period when there is no traffic, the observed "noise" in an outdoor environment is very different from an indoor or calibration environment, in which the noise spectrum is essentially white. Outdoors, even at night noise with transient periods and trends of small amplitudes is observed as seen in Figure 4.

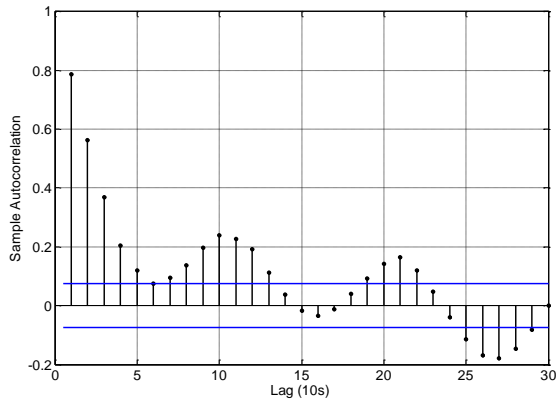

**Figure 7: Autocorrection of subsampled at 0.1Hz data.**

---

[1] The term "noise" is loosely defined here as the variation in the output electronic signal, which is not caused by immediate traffic conditions. This includes electronic noise, background pollution, pollution from other non-traffic sources, atmospheric changes and distance traffic.

The acf and pacf of 0.1Hz subsampled series suggest a AR(1) model for the data with AR parameter $\phi_1 = [0.77]$. This implies that $x_{t+1}$ is likely to be of the same sign as $x_t$, where the observed time series $X_t = \{x_1, x_2, \ldots x_n\}$ is defined as individual data samples collected in the experiment. However, this linear model does not fully describe the data characteristics. Firstly, local trends and periods can be observed in the data, but over longer time scales (>20 mins) the data appears to be stationary, similar to a nonlinear long memory process [7]. The nonlinearity in the data can be captured using the Delayed Residual Map (DRM) suggested in [8] by plotting the vectors $(x_t, r_{t+1})$, where $r_{t+1}$ is the residuals of the real value $x_{t+1}$ and the predicted value $\hat{x}_{t+1}$ using linear AR(1) model, and are defined by

$$r_t = x_t - a_0 - \sum_{i=1}^{m} a_i x_{t-i} \tag{1}$$

where $a_0, a_1, \ldots, a_m$ are the coefficients of the $AR(m)$ model obtained by minimizing the sums of squares of the residuals. DRM provides a simple estimate for the "lag-one nonlinearity" $f_1(x)$ in a time series model of the form,

$$x_{t+1} = f_1(x) + a_0 + \sum_{i=0}^{m-1} a_i x_{t-i} + n_t \tag{2}$$

where $a_0, a_1, \ldots, a_m$ are the parameters in (1) defining the linear part of the models and $n_t$ is white noise.

The values of $x_t$ (from the 0.1Hz subsampled series) are sorted into 7 equal bins and the mean of $r_{t+1}$ is plotted against the centre $x_t$ values (the choice of 7 is made for the purpose of easy visualisation) in Figure 8. More details of DRM and its application for analysing weather and epilepsy data can be found in [8].

Figure 8 depicts the DRM calculated from the 2 hours data of 9/2/2012 4-6pm. We observe the one-step ahead non-linearity of $r_{t+1}$ on the value $x_t$. With small $x_t$, the $r_{t+1}$ tends to be slightly positive, meaning that $x_{t+1}$ is more likely to increase. $r_{t+1}$ declines quickly and becomes negative with large $x_t$, indicating $x_{t+1}$ is likely to decrease. Note that majority of $x_t$ values are low, as depicted in the histogram in Figure 5 and Figure 6. The combination of these two characteristics can explain the long-term stationarity of the time series.

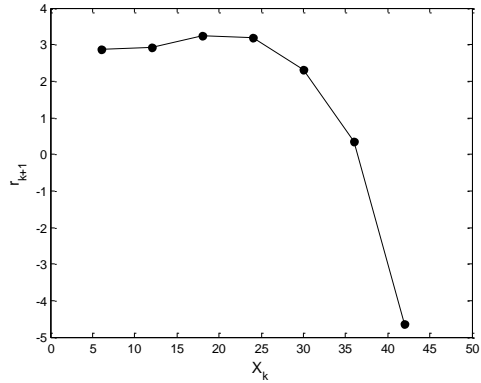

**Figure 8: Residual Delay Maps, with linear prediction with AR (1) model**

## 2.3 Underlying Dominant Frequency

Periodicities observed in Figure 7 can be related to the traffic flow caused by the sequence of traffic lights. Frequency analysis was performed on the subsample 0.2Hz data series, as shown in Figure 9. $1/f$ noise is observed in addition to a the major frequency component. The dominant frequency component varies with time of the day dependent on the traffic flow and is observed most clearly between the hours from 11am-7pm, as shown in Figure 10. Note that this frequency characteristic was particular to our experiments carried out on that day and location.

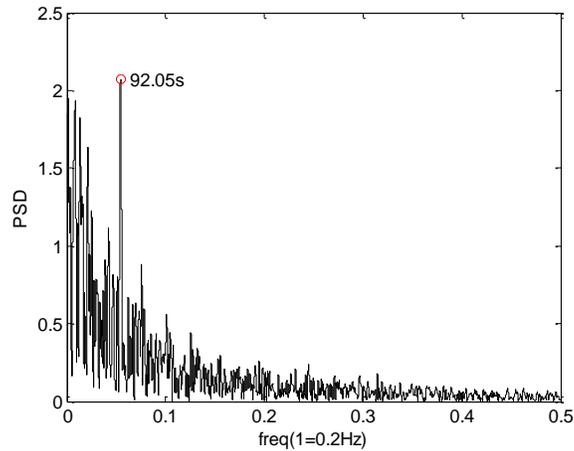

**Figure 9: Power spectral density (PSD) of subsampled 0.2Hz series.**

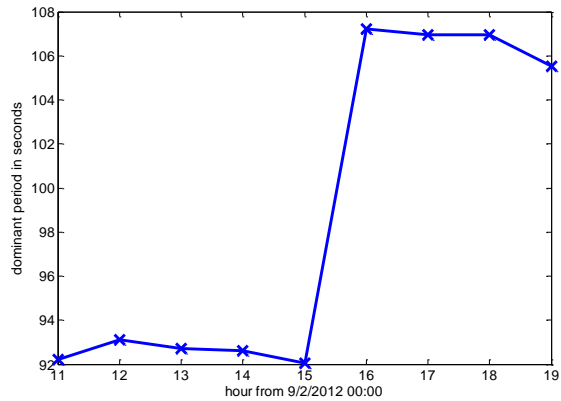

**Figure 10: Dominant frequency between 11am and 7pm on 9/2/2012**

## 3. CONCLUSION

Some interesting results are found in the data analysis performed on the pollution data, collected on 9-10/2/2012 in Hyderabad, India. We verified some of the previous findings in other literature including pollution data are log-normally distributed. Using Delayed Residual Maps (DRM), the residuals of linear model AR(1), and hence, the measured time series, appear to be nonlinear. Dominant frequencies are found during hours 11am-7pm with periods of 90 seconds and 110 seconds. Pollution data is log-normally distributed in the day time and night time data.